\documentclass[prl,groupedaddress,twocolumn]{revtex4-1}
\usepackage[dvips]{graphicx}
\usepackage{subfigure}
\usepackage{amsmath}
\usepackage{bm}
\usepackage{dcolumn}
\begin{document}

\title{Two-photon super bunching of thermal light via multiple two-photon-path interference}

\author{Peilong Hong}
\affiliation{The MOE Key Laboratory of Weak Light Nonlinear Photonics and School of Physics, Nankai University, Tianjin 300457, China}

\author{Jianbin Liu}
\affiliation{The MOE Key Laboratory of Weak Light Nonlinear Photonics and School of Physics, Nankai University, Tianjin 300457, China}

\author{Guoquan Zhang}
\email{zhanggq@nankai.edu.cn}
\affiliation{The MOE Key Laboratory of Weak Light Nonlinear Photonics and School of Physics, Nankai University, Tianjin 300457, China}

\date{\today}

\begin{abstract}

We propose a novel scheme to achieve two-photon super bunching of thermal light through multiple two-photon-path interference, in which two mutually first-order incoherent optical channels are introduced by inserting a modified Michelson interferometer into a traditional two-photon HBT interferometer, and the bunching peak-to-background ratio can reach 3 theoretically. Experimentally, the super bunching peak-to-background ratio was measured to be 2.4,  much larger than the ratio 1.7 measured with the same thermal source in a traditional HBT interferometer. The peak-to-background ratio of two-photon super bunching of thermal light can be increased up to $2\times1.5^n$ by inserting cascadingly $n$ pairs of mutually first-order incoherent optical channels into the traditional two-photon HBT interferometer. The two-photon super bunching of thermal light should be of great significance in improving the visibility of classical ghost imaging.

\end{abstract}

\maketitle

Two-photon bunching of thermal light was first observed by Hanbury Brown and Twiss (HBT) in 1956~\cite{HBT}, where the maximum  bunching peak-to-background ratio for thermal light is 2. One of the prospective applications of two-photon bunching effect is ghost imaging, and extensive studies have been carried out by using various light sources~\cite{PITTMAN95,BOYD02,GATTIPRL04,WUOL05,ZHUPRE05,SHIHPRL06}. However, the imaging visibility of the classical ghost imaging, especially with complicated imaging structures, is relatively low based on the traditional two-photon HBT interference of thermal light. To overcome this limitation, spatial super bunching of thermal light with a bunching peak-to-background ratio larger than 2 has attracted a lot of interests, and great progresses have been made recently to enhance the visibility of classical ghost imaging~\cite{HANPRA07, BOYDOL09, WUOL10, SHIHPRA10,WANGAPL08,LIUPRA09}. It was demonstrated that the visibility of classical ghost imaging is improved by employing $n$th-order ($n>2$) coherence of thermal light, in which the bunching peak-to-background ratio reaches $n!$~ \cite{SHIHPRA10,WANGAPL08,LIUPRA09}. For the two-photon case, super bunching effect was observed with laser beam scattered by a dynamic deep random phase screen with non-Gaussian statistics~\cite{JPA75t,JPA75e}. Recently, two-photon super bunching effect was also observed for thermal-like photons with attractive interaction between photons propagating in a nonlinear medium with a focusing nonlinearity~\cite{np10}. In this Letter, we report on the two-photon super bunching effect of thermal light by employing multiple two-photon-path interference, which would be of great significance in improving the visibility of classical ghost imaging.

\begin{figure}[!htb]
\centering
\subfigure{\label{mul0}
\includegraphics[width=0.13\textwidth]{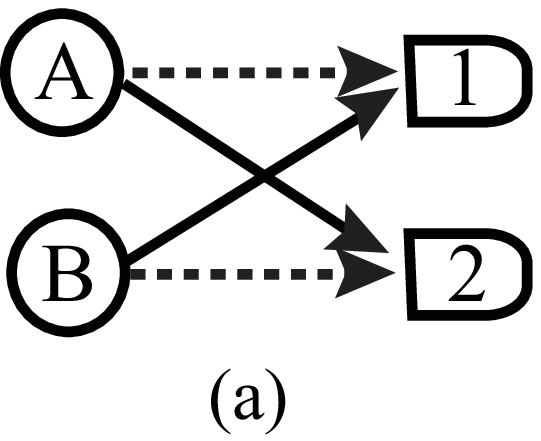}}
\subfigure{\label{mul1}
\includegraphics[width=0.21\textwidth]{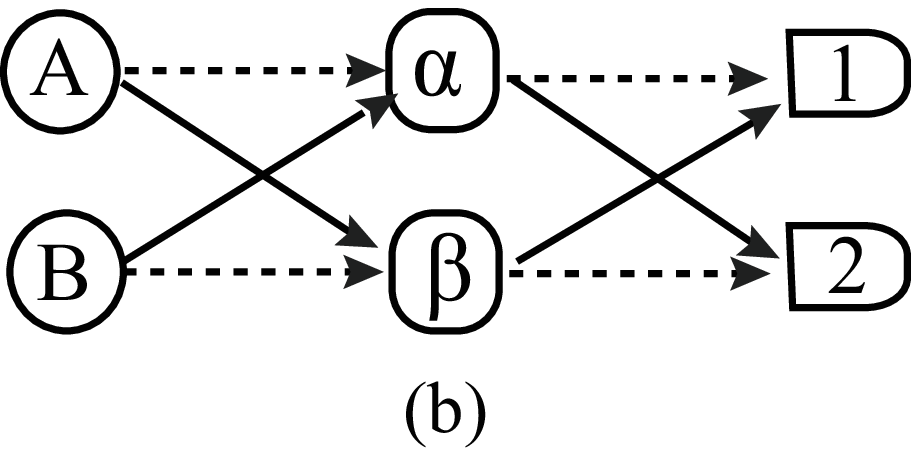}}
\caption{Schematic diagrams for the two-photon paths of thermal light in a traditional HBT interferometer (a) and those in our interferometer with two mutually first-order incoherent intermediate channels $\alpha$ and $\beta$ (b), respectively. Here $A$ and $B$ are two different photons to trigger a coincidence count,  $1$ and $2$ are two single-photon detectors, respectively. }\label{mu}
\end{figure}

Figure~\ref{mul0} shows a schematic diagram for a traditional two-photon HBT interferometer. It is well known that the key of the two-photon bunching effect is the existence of two different but indistinguishable two-photon paths to trigger a coincidence count, i.e., $A1B2$ and $A2B1$, where $A$ and $B$ denote two different photons, and 1 and 2 denote two different detectors, respectively \cite{LIUPRA09,SHIHEL04,SHIHPRL06,MANDELRMP99,GLAUBERPR63}. Generally, in the $n$th-order coherence with $n$ detectors, the number of different but indistinguishable $n$-photon paths to trigger a $n$-photon coincidence count is $n!$, which results in a bunching peak-to-background ratio of $n!$~\cite{LIUPRA09, SHIHPRA10}. Different from the case in high-order coherence, here we propose a scheme to achieve two-photon super bunching of thermal light by increasing the number of different but indistinguishable two-photon paths.

Figure~\ref{mul1} shows a simple scheme which provides multiple different but indistinguishable two-photon paths by inserting a pair of intermediate optical channels $\alpha$ and $\beta$ into a traditional HBT interferometer. The two intermediate channels are incoherent with each other in single photon domain, i.e., there is no first-order interference in the interferometer. This can be realized, for example, by adding a time-variable random phase $\phi(t)$ on one intermediate channel. It is evident that the scheme reduces to a traditional HBT interferometer when one of the intermediate channels is blocked.

In the scheme in Fig.~\ref{mul1}, the two-photon paths that can trigger a coincidence count are

  $A\alpha1B\alpha2$, $A\alpha2B\alpha1$, $A\beta1B\beta2$, $A\beta2B\beta1$,

  $A\alpha1B\beta2$, $A\alpha2B\beta1$, $A\beta1B\alpha2$,  $A\beta2B\alpha1$,

\noindent respectively. Note that the two pairs of two-photon paths $A\alpha1B\alpha2$, $A\alpha2B\alpha1$ and $A\beta1B\beta2$, $A\beta2B\beta1$ in the first row are the contributions from two different traditional HBT interferometers. The four two-photon paths in the second row are different but indistinguishable. We will show that, it is these four different but indistinguishable two-photon paths that result in a two-photon super bunching of thermal light.

Mathematically, the second-order coherence function of thermal light is~\cite{GLAUBERPR63}
  \begin{equation} \label{glauber}
    \begin{split}
     G^{(2)}({\bf{r}}_{1},t_{1};{\bf{r}}_{2},t_{2})=&Tr\big\{\widehat{\rho}\widehat{E}^{(-)}({\bf{r}}_{1},t_{1})\widehat{E}^{(-)}({\bf{r}}_{2},t_{2})\\
     \times &\widehat{E}^{(+)}({\bf{r}}_{2},t_{2})\widehat{E}^{(+)}({\bf{r}}_{1},t_{1})\big\} \, ,
    \end{split}
  \end{equation}
where $\widehat{\rho}=\sum_{\{n\}}P_{\{n\}}\big|\{n\}\big\rangle\big\langle\{n\}\big|$ is the density matrix of thermal light with $P_{\{n\}}$ being the probability of the field in state $\big|\{n\}\big\rangle$~\cite{LIUPRA09,LOUDON00}, $\widehat{E}^{(+)}({\bf{r}}_{j},t_{j})$ and $\widehat{E}^{(-)}({\bf{r}}_{j},t_{j})$ $(j=1,2)$ are the positive and negative frequency parts of the field at the
space-time coordinate $({\bf{r}}_{j},t_{j})$, respectively. The field operator $\widehat{E}^{(+)}({\bf{r}}_{j},t_{j})$ is expressed as \cite{LIUPRA09,RUBINPRA96}
  \begin{equation}\label{field}
 \widehat{E}^{(+)}(\vec{\rho}_{j},z_{j},t_{j})=\int
      g(\vec{\rho}_{j},z_j;\vec{\kappa},\omega)
      a(\vec{\kappa},\omega) e^{-i\omega t_j} d \omega d \vec{\kappa} \, ,
   \end{equation}
where $g(\vec{\rho}_{j},z_j;\vec{\kappa},\omega)$ is the Green's function describing the light in mode $(\vec{\kappa},\omega)$ propagating to space-time coordinate $({\bf{r}}_{j},t_{j})$, and it can be expressed as
  \begin{equation} \label{green}
    \begin{split}
      g(\vec{\rho}_{j},z_j;\vec{\kappa},\omega)&=\frac{-i\omega}{2\pi c}\frac{e^{ikz_j}}{z_j}
       \int d\vec{\rho}_s  A(\vec{\rho}_s)e^{-i\vec{\kappa}\cdot \vec{\rho}_s}\\
      &\times \psi(\frac{\omega}{c z_j},\vec{\rho}_j-\vec{\rho}_s) \,,
    \end{split}
  \end{equation}

\noindent
where $\psi(\frac{\omega}{c z_j},\vec{\rho}_j-\vec{\rho}_s)=e^{i\frac
{\omega}{2cz_j}|\vec{\rho}_j-\vec{\rho}_s|^2}$.
 $\vec{\rho}_{j(s)}$ and $z_{j(s)}$ are the transverse and longitudinal parts of the spatial coordinates of the $j$th detector and the source $s$, respectively.  $ A(\vec{\rho}_s)$ is the amplitude of thermal light in mode
$(\vec{\kappa}, \omega)$.

The coherence function of thermal light in a traditional HBT interferometer can be deduced as \cite{LIUPRA09}
  \begin{equation} \label{glauber-hbt}
     \begin{split}
           G_{HBT}^{(2)}({\bf{r}}_{1},{\bf{r}}_{2}) &=\sum_{\cdots,n(\vec{\kappa})\ge 1,
           n(\vec{\kappa'})\ge 1,\cdots} n(\vec{\kappa})n(\vec{\kappa'}) P_{\{n\}}\\
           &\times \int d\vec{\kappa} d\vec{\kappa'}
           \Big | \frac{1}{\sqrt{2}}\big[ g(\vec{\rho}_{1},z_1;\vec{\kappa})g(\vec{\rho}_{2},z_2;\vec{\kappa'})\\
           &+g(\vec{\rho}_{2},z_2;\vec{\kappa})g(\vec{\rho}_{1},z_1;\vec{\kappa'})\big ]\Big|^2 \, ,
     \end{split}
   \end{equation}
where the time-related variables have been dropped since we are only interested in the spatial part and treat the light as a single-frequency one. Taking the paraxial approximation in one dimension and assuming that $A(\vec{\rho}_s)$ is a constant and $z_1=z_2=z$, we get the normalized second-order coherence function of thermal light in a traditional HBT interferometer
\begin{equation}\label{HBT}
      g_{HBT}^{(2)}(x_{1},x_{2})=1+sinc^2\big[\frac{kR(x_1-x_2)}{2z}\big] \, ,
\end{equation}
where $k$ is the magnitude of the wave vector of thermal light, and $R$ is the cross-section size of the source $s$.

In the scheme shown in Fig.~\ref{mul1}, the Green's function can be expressed as
\begin{equation}\label{green2}
     g(\vec{\rho}_{j},z_j;\vec{\kappa})=g_\alpha
     (\vec{\rho}_{j},z_j;\vec{\kappa})+g_\beta (\vec{\rho}_{j},z_j;\vec{\kappa}) e^{i\phi(t)}\, ,
\end{equation}
where $g_\alpha (\vec{\rho}_{j},z_j;\vec{\kappa})$ and $g_\beta(\vec{\rho}_{j},z_j;\vec{\kappa})e^{i\phi(t)}$ are the Green's functions associated with the photon propagating through the channels $\alpha$ and $\beta$, respectively, to the $j$th detector. Here we employ the time-variable random phase $\phi(t)$ in channel $\beta$ to eliminate the first-order interference pattern between the two channels, and it satisfies
  \begin{equation}
    \big\langle e^{i\phi(t)}\big\rangle_t=0\, ,
\end{equation}
where $\langle\cdots\rangle_t$ denotes time average. Replacing the Green's function in Eq.(\ref{glauber-hbt}) by Eq. (\ref{green2}), one gets the spatial correlation function of thermal light in the scheme in Fig.~\ref{mul1}
    \begin{equation} \label{glauber2}
     \begin{split}
        &G^{(2)}({\bf{r}}_{1},{\bf{r}}_{2})=\\
          &\sum_{\cdots,n(\vec{\kappa})\ge 1,n(\vec{\kappa'})\ge 1,\cdots} n(\vec{\kappa})n(\vec{\kappa'}) P_{\{n\}}  \int d\vec{\kappa} d\vec{\kappa'}\frac{1}{2}\\
         &\times \Big[ \big | g_\alpha(\vec{\rho}_{1},z_1;\vec{\kappa})g_\alpha(\vec{\rho}_{2},z_2;\vec{\kappa'})
         +g_\alpha(\vec{\rho}_{2},z_2;\vec{\kappa})g_\alpha(\vec{\rho}_{1},z_1;\vec{\kappa'})\big|^2 \\
         &+ \big | g_\beta(\vec{\rho}_{1},z_1;\vec{\kappa})g_\beta(\vec{\rho}_{2},z_2;\vec{\kappa'})
         +g_\beta(\vec{\rho}_{2},z_2;\vec{\kappa})g_\beta(\vec{\rho}_{1},z_1;\vec{\kappa'})\big|^2 \\
         &+\big | g_\alpha(\vec{\rho}_{1},z_1;\vec{\kappa})g_\beta(\vec{\rho}_{2},z_2;\vec{\kappa'})
         +g_\alpha(\vec{\rho}_{2},z_2;\vec{\kappa})g_\beta(\vec{\rho}_{1},z_1;\vec{\kappa'}) \\
         &+g_\beta(\vec{\rho}_{1},z_1;\vec{\kappa})g_\alpha(\vec{\rho}_{2},z_2;\vec{\kappa'})
         +g_\beta(\vec{\rho}_{2},z_2;\vec{\kappa})g_\alpha(\vec{\rho}_{1},z_1;\vec{\kappa'})\big|^2\Big]\, .
    \end{split}
  \end{equation}
Again, by employing the paraxial approximation in one dimension and assuming that $A(\vec{\rho}_s)$ is a constant and $z_1=z_2=z$, the normalized second-order coherence function of thermal light in the scheme in Fig.~\ref{mul1} can be deduced from Eq. (\ref{glauber2}) as
  \begin{equation}\label{glauber-fringe}
    \begin{split}
      g^{(2)}(x_{1},x_{2})&=\big[1+sinc^2\big(\frac{kR(x_1-x_2)}{2z}\big)\big]\\
      &\times \big[1+\frac{1}{2}cos\big(k\theta(x_1-x_2)\big)\big]\, ,
     \end{split}
  \end{equation}
where $\theta$ is the crossing angle between the two optical paths associated with a photon propagating through the two channels $\alpha$ and $\beta$ to a detector. Eq.~(\ref{glauber-fringe}) shows a two-photon interference pattern with a period of $2\pi/k\theta$ and a peak amplitude of 3, while its amplitude envelope is modulated by the traditional two-photon bunching curve of thermal light.

\begin{figure}[!htb]
    \centering
    \includegraphics[width=0.4\textwidth]{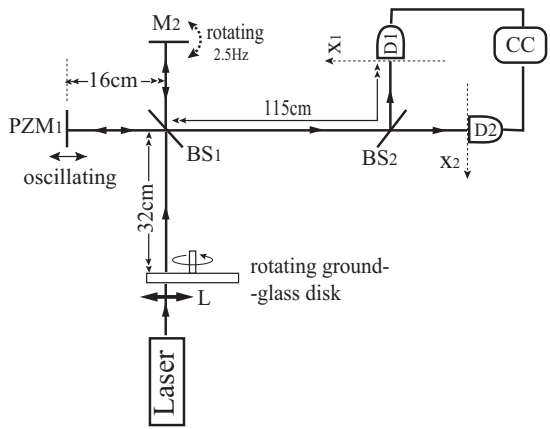}
    \caption{Experimental setup scheme. L: lens, PZM1: piezo-electric mirror, M2: mirror, BS1 and BS2: beam splitters, D1 and D2: single-photon detectors, CC: two-photon coincidence counting system.
    \label{exp}}
\end{figure}

\begin{figure}[!htb]
    \centering
    \includegraphics[width=0.38\textwidth]{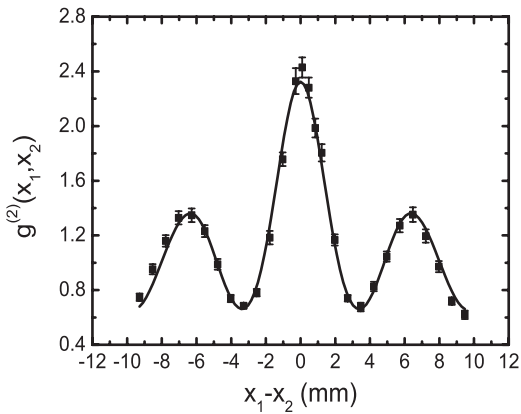}
    \caption{Two-photon interference fringe induced by the superposition of multiple two-photon paths. The solid squares are the experimental data, and the curve is a theoretical fit employing Eq.~(\ref{glauber-fringe}) with $z=179$ cm, $R=356$ $\mu$m and $\theta=0.007^{o}$. \label{fringe}}
\end{figure}

Experimentally, we observed such a two-photon interference pattern induced by the superposition of multiple two-photon paths, where the two intermediate channels were introduced by inserting a modified Michelson interferometer into a traditional two-photon HBT interferometer, as shown in Fig.~\ref{exp}. In the experiments, the thermal source, with a cross-section size $R=347$ $\mu$m, was simulated by a focused single frequency 780-nm laser beam  scattering from a rotating ground glass disk~\cite{SPILLERAJP64}. The thermal light was then launched into the input port of a Michelson interferometer, and the two arms of the Michelson interferometer served as the two optical channels $\alpha$ and $\beta$. The end mirror PZM1 on one arm of the Michelson interferometer (channel $\beta$) was a piezo-electric mirror, and it oscillated along the light propagation direction to provide a time-variable phase $\phi(t)$ in such a way that the first-order coherence between two optical channels was eliminated. The output of the Michelson interferometer was then split into two beams which were detected by two single-photon detectors. The spatial correlation was performed with a two-photon coincidence counting system with a collection time of 300 seconds. Figure~\ref{fringe} shows the observed two-photon interference fringe at $\theta=0.007^{o}$ with a peak amplitude of 2.4, in which the solid squares are the experimental data and the solid curve is a theoretical fit by employing Eq.~(\ref{glauber-fringe}). Good agreement is observed between experimental measurements and theoretical fit.

To suppress the side lobes in Fig.~\ref{fringe}, we rotated the mirror M2 on the other arm (channel $\alpha$) of the Michelson interferometer repeatedly at a frequency of 2.5 Hz which makes the crossing angle $\theta$ scan within $[-\theta_0/2, \theta_0/2]$ at a speed of $0.11^{o}/s$, where $\theta_0 (=0.022^{o})$ was the full angular scanning range. Theoretically, such an angular scanning leads to an average over $\theta$ in Eq.~(\ref{glauber-fringe}). Note that the fringe period $2\pi/k\theta$ is dependent on $\theta$, and the fringes with different $\theta$ are always in phase at $x_1-x_2=0$, but gradually become out of phase with the increase in the distance $|x_1-x_2|$. Therefore, the average over $\theta$ smears out the side lobes, and one gets the normalized two-photon correlation function as
 \begin{equation}  \label{glauber-bunching}
   \begin{split}
        g^{(2)}(x_{1},x_{2})=&\big[1+sinc^2\big(\frac{kR(x_1-x_2)}{2z}\big)\big]\\
        &\times \Big \langle 1+\frac{1}{2}cos\big(k\theta(t)(x_1-x_2)\big)\Big\rangle_t\\
        =&\big[1+sinc^2\big(\frac{kR(x_1-x_2)}{2z}\big)\big]\\
        &\times \Big[1+\frac{1}{2} sinc\big(\frac{k\theta_0(x_1-x_2)}{2}\big)\Big]\, .
   \end{split}
\end{equation}
\noindent It is evident that the bunching peak-to-background ratio is 3 in this case, much larger than 2 in the traditional two-photon HBT interferometer. Further more, the full width at half maximum (FWHM) of the bunching curve is determined by the angular scanning range $\theta_0$ and the ratio $R/z$, In the case when $\theta_0>R/z$, the FWHM of the super bunching curve can be narrowed as compared to that of the traditional two-photon bunching curve.

\begin{figure}[!htb]
    \centering
    \includegraphics[width=0.38\textwidth]{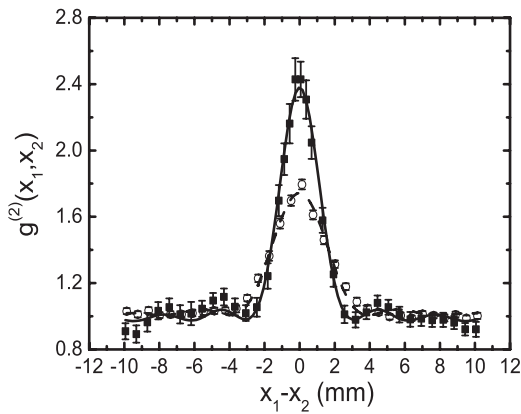}
    \caption{Two-photon bunching (empty circles and dashed curve) and super bunching (solid squares and solid curve) of thermal light. The circles and squares are the experimental data, and the dashed and solid curves are the theoretical fits by employing Eqs.~(\ref{HBT}) and (\ref{glauber-bunching}), respectively, with $z=179$ cm, $R=356$ $\mu$m and $\theta_{0}=0.026^{o}$. \label{bunching}}
\end{figure}

Figure~\ref{bunching} shows the measured two-photon super bunching effect (solid squares) of thermal light with $\theta_0=0.022^{o}$, together with the results for the two-photon bunching effect (empty circles) in the traditional HBT interferometer with the same thermal light source. The solid and the dashed curves are the theoretical fits using Eqs.~(\ref{glauber-bunching}) and (\ref{HBT}) for the super bunching and traditional bunching cases, respectively, which describe the experimental results very well in both cases. The super bunching peak-to-background ratio was measured to be 2.4, much larger than the ratio 1.7 measured for the traditional HBT bunching effect employing the same thermal source. Thus, we have demonstrated the two-photon super bunching effect of thermal light via multiple two-photon-path interference.

\begin{figure}[!htb]
    \centering
    \includegraphics[width=0.4\textwidth]{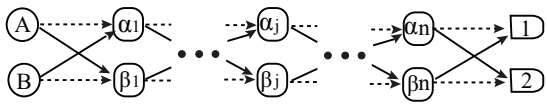}
    \caption{Schematic diagram for multiple two-photon paths of thermal light when $n$ pairs of intermediate optical channels are inserted. Here $\alpha_j$ and $\beta_j$ are the $j$th-pair of intermediate channels, other symbols are the same as those in Fig.~\ref{mu}.}\label{mul}
\end{figure}

It is easy to figure out that the number of different but indistinguishable two-photon paths can be further increased by inserting cascadingly $n$ pairs of intermediate optical channels in the traditional two-photon HBT interferometer, as shown schematically in Fig.~\ref{mul}, where $\alpha_j$ and $\beta_j$ represent the $j$th-pair of intermediate optical channels. In this case, by employing the same procedures as those from Eq.~(\ref{green2}) to Eq.~(\ref{glauber-bunching}),
one gets
\begin{equation}
\begin{split} \label{FP}
       g^{(2)}(x_{1},&x_{2})=\big[1+sinc^2\big(\frac{kR(x_1-x_2)}{2z}\big)\big]\\
       &\times \prod_{j=1}^{n}\Big[1+ \frac{1}{2} sinc\big(\frac{k\theta_{j0}(x_1-x_2)}{2}\big)\Big]\, ,
\end{split}
\end{equation}
\noindent where $\theta_{j0}$ is the full angular scanning range associated with the $j$th-pair of intermediate optical channels.

One notes that the two-photon super bunching peak-to-background ratio of thermal light in Eq.~(\ref{FP}) can reach $2\times1.5^n$, which is much larger than that in traditional two-photon HBT interferometer. Such a super bunching effect is surely the result of the superposition of multiple different but indistinguishable two-photon paths introduced by inserting $n$ pairs of intermediate optical channels. This is very similar to the case for the multiple single-photon-path interference in first-order coherence, in which the superposition of multiple different but indistinguishable single-photon paths results in an enhancement and sharpen of the first-order coherence peak, as is the case for a traditional one-photon grating~\cite{BROOKER03}. Therefore, the scheme we designed can be viewed as a prototype of two-photon grating.

In summary, we have demonstrated the two-photon super bunching of thermal light by means of superposition of multiple different but indistinguishable two-photon paths in a modified two-photon interferometer. By inserting $n$ pairs of intermediate optical channels in a traditional two-photon HBT interferometer, the super bunching peak-to-background ratio of thermal light can be increased up to $2\times 1.5^n$. The super bunching peak-to-background ratio was measured to be 2.4 experimentally when a pair of intermediate optical channels were introduced, while the bunching peak-to-background ratio of the same thermal source was measured to be 1.7 in a traditional two-photon HBT interferometer. The observed two-photon super bunching effect of thermal light should be useful to improve the visibility of classical ghost imaging.

\begin{acknowledgments}
This project is supported by the MOE Cultivation Fund of the Key Scientific and Technical Innovation Project (708022), the NSFC (90922030, 10804054, 10904077),  the 973 programs (2007CB307002, 2011CB922003), the 111 project (B07013), and the Fundamental Research Funds for the Central Universities.
\end{acknowledgments}


\begin{thebibliography}{99}
\bibitem{HBT} R. Hanbury Brown and R. Q. Twiss, Nature(London) \textbf{177}, 27 (1956); \textbf{178}, 1046 (1956).

\bibitem{PITTMAN95} T. B. Pittman, Y. H. Shih, D. V. Strekalov, and A. V. Sergienko, Phys. Rev. A \textbf{52}, R3429 (1995).

\bibitem{BOYD02} R. S. Bennink, S. J. Bentley and R. W. Boyd, Phys. Rev. Lett. \textbf{89}, 113601 (2002).

\bibitem{GATTIPRL04} A. Gatti, E. Brambilla, M. Bache, and L. A. Lugiato, Phys. Rev. Lett. \textbf{93}, 093602 (2004); Phys. Rev. A \textbf{70}, 013802 (2004).

\bibitem{SHIHPRL06}A. Valencia, G. Scarcelli, M. D'Angelo, and Y. Shih, Phys. Rev. Lett. \textbf{94}, 063601 (2005); G. Scarcelli, V. Berardi, and Y. Shih, Phys. Rev. Lett. \textbf{96}, 063602 (2006).

\bibitem{WUOL05}  D. Zhang, Y. H. Zhai, L. A. Wu, and X. H. Chen, Opt. Lett. \textbf{30}, 2354 (2005).

\bibitem{ZHUPRE05}Y. J. Cai and S. Y. Zhu, Phys. Rev. E \textbf{71}, 056607(2005).

\bibitem{HANPRA07}Y. Bai and S. S. Han, Phys. Rev. A \textbf{76}, 043828 (2007).

\bibitem{BOYDOL09}K. W. C. Chan, M. N. O'Sullivan, and R. W. Boyd, Opt. Lett. \textbf{34}, 3343 (2009).

\bibitem{WUOL10}X. Chen, I. N. Agafonov, K. Luo, Q. Liu, R. Xian, M. V. Chekhova, and L. Wu, Opt. Lett. \textbf{35}, 1166 (2010).

\bibitem{SHIHPRA10}Y. Zhou,  J. Simon, J. Liu, and Y. Shih, Phys. Rev. A \textbf{81}, 043831 (2010).

\bibitem{WANGAPL08}D. Cao, J. Xiong, S. Zhang, L. Lin, L. Gao, and K. Wang, Appl. Phys. Lett. \textbf{92}, 201102 (2008).

\bibitem{LIUPRA09} J. Liu and Y. Shih, Phys. Rev. A \textbf{79}, 023819 (2009).

\bibitem{JPA75t} E. Jakeman and P. N. Pusey, J. Phys. A \textbf{8}, 369 (1975).

\bibitem{JPA75e} P.  N. Pusey and E. Jakeman, J. Phys. A \textbf{8}, 392 (1975).

\bibitem{np10}Y. Bromberg, Y. Lahini, E. Small, and Y. Silberberg, Nature Photonics \textbf{4}, 721 (2010).

\bibitem{MANDELRMP99}L. Mandel, Rev. Mod. Phys. \textbf{71}, S274 (1999).

\bibitem{SHIHEL04}G. Scarcelli, A. Valencia and Y. Shih, Europhys. Lett. \textbf{68}, 618 (2004).

\bibitem{GLAUBERPR63} R. J. Glauber, Phys. Rev. \textbf{130}, 2529 (1963); Phys. Rev. \textbf{131}, 2766 (1963).

\bibitem{LOUDON00} R. Loudon, \emph{The Quantum Theory of Light} (Oxford University Press, 2000).

\bibitem{RUBINPRA96} M. H. Rubin, Phys. Rev. A \textbf{54}, 5349 (1996).

\bibitem{SPILLERAJP64} W. Martienssen and E. Spiller, Am. J. Phys. \textbf{32}, 919 (1964).

\bibitem{BROOKER03} G. Brooker, \emph{Modern Classical Optics} (Oxford University Press, 2003).


\end{thebibliography}
\end{document}